\documentclass{article}
\usepackage{graphicx}
\usepackage{amssymb}
\usepackage{amsmath}
\usepackage{amsfonts}
\usepackage{hyperref}
\usepackage{chapterbib}
\usepackage{geometry}
\usepackage{multirow}
\usepackage{subcaption}
\usepackage{epstopdf}
\graphicspath{{}}
\DeclareMathOperator*{\argmax}{arg max}

\begin{document}

\title{Classification of Audio Segments in Call Center Recordings using Convolutional Recurrent Neural Networks
}

\author{Şükrü Ozan, Ph.D. \href{https://orcid.org/0000-0002-3227-348X}{\includegraphics[scale=0.075]{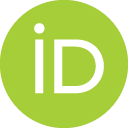}}\\sukruozan@gmail.com }


\date{June 2021}

\maketitle

\begin{abstract}
Detailed statistical analysis of call center recordings is critical in the customer relationship management point of view. With the recent advances in artificial intelligence, many tasks regarding the calculation of call statistics are now performed automatically. This work proposes a neural network framework where the aim is to correctly identify audio segments and classify them as either customer or agent sections. Accurately identifying these sections gives a fair metric for evaluating agents' performances. We inherited the convolutional recurrent neural network (CRNN) architecture commonly used for such problems as music genre classification. We also tested the same architecture's performance, where the previous class information and the gender information of speakers are also added to the training data labels. We saw that CRNN could generalize the training data and perform well on validation data for this problem with and without the gender information. Moreover, even the training was performed using Turkish speech samples; the trained network was proven to achieve high accuracy for call center recordings in other languages like German and English. 

\end{abstract}

{\bf Keywords:} Convolutional recurrent neural networks, Long short term memory, Call center recordings, Audio classification.

\section{Introduction}\label{sec:intro}

Telephony has been an essential tool used by companies for marketing and customer satisfaction. Moreover, if the services or products already meet customer needs, telephone calls guarantee customer satisfaction and loyalty for any company \cite{Feinberg:2000}. A thorough analysis of telephone calls and improving overall call performance is the most critical job for quality managers of companies that use telephony as a significant component of their customer relationship management (CRM) processes. Although it seems that the call center agents may not be replaced by artificial intelligence (AI) soon, the majority of call center quality managers seem to be so, since most of both quantitative and qualitative analysis tasks can now be performed by artificial intelligence. 

In this study, we propose a method to analyze phone calls between a call agent and a customer to classify a given speech segment as belonging to either a customer or an agent.  As of being another deep supervised learning procedure, also this solution method needs numerous labeled training data. Since we had no readily available training data,  we also propose a convolutional neural network (CNN) based procedure that can be used to produce labeled data for the training. The corresponding procedure will further be explained in the following sections. 

Big tech companies like Google, Amazon, and Facebook are now leading AI research. It is not unusual to meet a new product deployed by these companies, which solves a critical AI problem. However, it is still not affordable for the vast majority of small and medium-sized enterprises (SMEs) to use them in their daily business process cycles. AdresGezgini Inc. (from now on referred to as the company ) is an example located in Turkey. It is an internet ads and software development company serving for the last 13 years. The company mostly uses telephony for its CRM processes. For the last six years, the company manages its call operations by using a third-party call center software that keeps the basic call statistics and all the call recordings.  However, call center software cannot measure more detailed statistics like individual speaker durations. 
The manuscript is organized as follows: We start with review of previous studies that either directly or indirectly inspired us in this study. Then we give the details of the proposed method. Creating an accurate training data set was the most challenging and time-consuming process; hence, we give as much detail as possible about the training data. Then we explain the proposed convolutional recurrent neural network (CRNN) model architecture and its training process. The performance of the model on both training and validation data is reported in the results section. We spared another subsection for an extension of the model. Even if we primarily aim  to classify a given audio segment as either customer or agent, we also have gender information for each data. Hence, we increased the number of target classes to four, namely female customer, male customer, female agent, and male agent. We also report the results in the results section. We conclude the manuscript with concluding remarks and our insights on relevant future studies.

\section{Related Work}\label{sec:relatedwork}

Our primary intuition was that the agent speech and customer speeches have specific rhythm and style. Hence, we hypothesize that identifying agents and customers from given audio segments resemble the music genre classification problem. While building up our solution method, we made an analogy with this problem. The music genre prediction problem has been drawing significant attention in audio related literature for the last few decades.

The idea of using Mel-frequency cepstral coefficients (MFCCs) in audio processing research has been much appreciated since 1980 \cite{Davis:1980}. It is still a vital feature extraction method for state of the art deep learning methods to solve various audio processing problems. Likewise, we use MFCCs to extract features to be used as training data. In the 2000s, to perform music genre classification, radial basis functions (RBFs) with MFCC and STFT features were proven to achieve close to human-level performance (HLP)  \cite{Turnbull:2005}. In \cite{Meng:2007}, temporal information integration of audio features is performed using multivariate autoregressive (MAR) features, and HLP comparable accuracies were achieved. On the contrary, these previous studies needed carefully tailored handcrafted features. 

With the recent advances in machine learning (ML) deep learning (DL), audio-related problems are now solved with neural networks if there is a sufficient number of training data samples. Since audio signals have implicit temporal information, using recurrent neural networks (RNNs)  gives better results in audio-related problems like speaker verification \cite{Wan:2018}. Moreover, CRNNs perform well in some problems since they can learn deep features and relate temporal information embedded in sequential data like audio signals \cite{Keren:2016}. 

In \cite{Choi:2017}, music tagging is performed using CRNNs, and it is reported that it outperforms three different CNN architectures built with convolutional and fully connected layers. A similar CRNN architecture is used for music artist classification in \cite{Nasrullah:2019}. In this study, we use a slightly modified version of the architecture proposed in  \cite{Choi:2017}. In the recurrent layer, we tested both gated recurrent unit (GRU) and  long short term memory (LSTM)  units. The structural details and differences of these units can be seen in \cite{Goodfellow:2016}. The effect of using either GRU or LSTM on performance is reported in the Results section. 

This study needs preprocessing of the files to separate customer and agent speeches since the call center conversations are recorded and compressed as single-channel .gsm (i.e., GSM 6.10 prI-ETS 300 036 13kbit/s Standard) files with 8 kHz sampling frequency for disk usage optimization.  A thorough analysis of some state-of-the-art methods for speech separation in machine learning literature can be found in \cite{Wang:2018}. In \cite{Luo:2019}, an end-to-end convolutional time-domain audio separation framework shows significant performance in speaker separation for single-channel audio.

On the other hand, since the call center recordings are dialogues, i.e., the speaker voices do not overlap, speaker diarization can also be effectively used to solve the "who spoke when" problem for them. In \cite{Zhang:2019}, Google researchers proposed a  speaker diarization system that uses generalized end to end (GE2E) loss, which is proposed initially for speaker verification in \cite{Wan:2018}.  The GE2E loss utilizes d-vectors \cite{Heigold:2016}, which are embeddings generated using LSTM based deep neural networks (DNNs). 

\begin{figure*}[t!]
\begin{subfigure}{.50\textwidth}
  \centering
  \includegraphics[width=\linewidth]{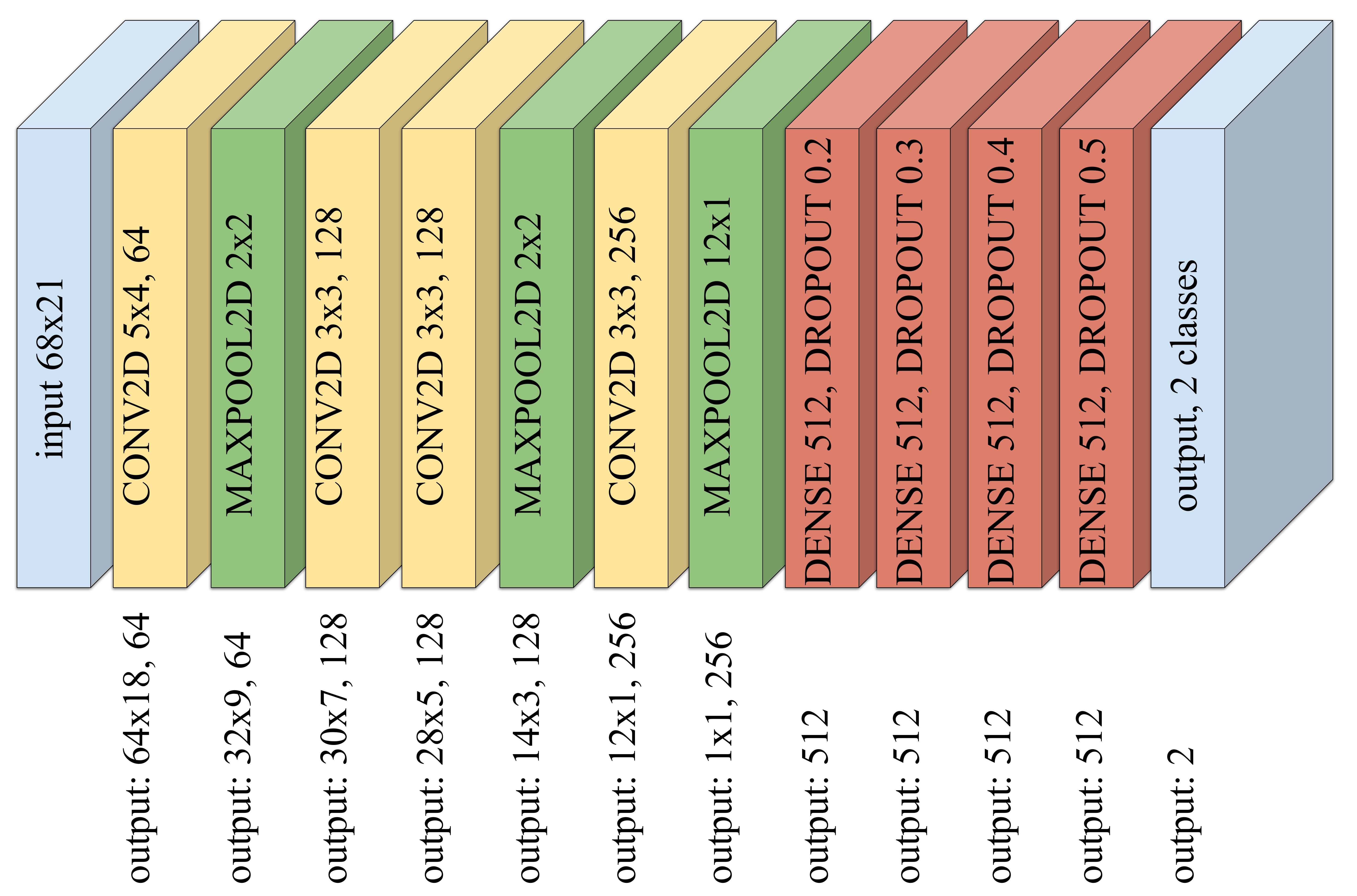}  
  \caption{ Speech-music classification architecture }
  \label{fig:inaMusic}
\end{subfigure}
\begin{subfigure}{.50\textwidth}
  \centering
  \includegraphics[width=\linewidth]{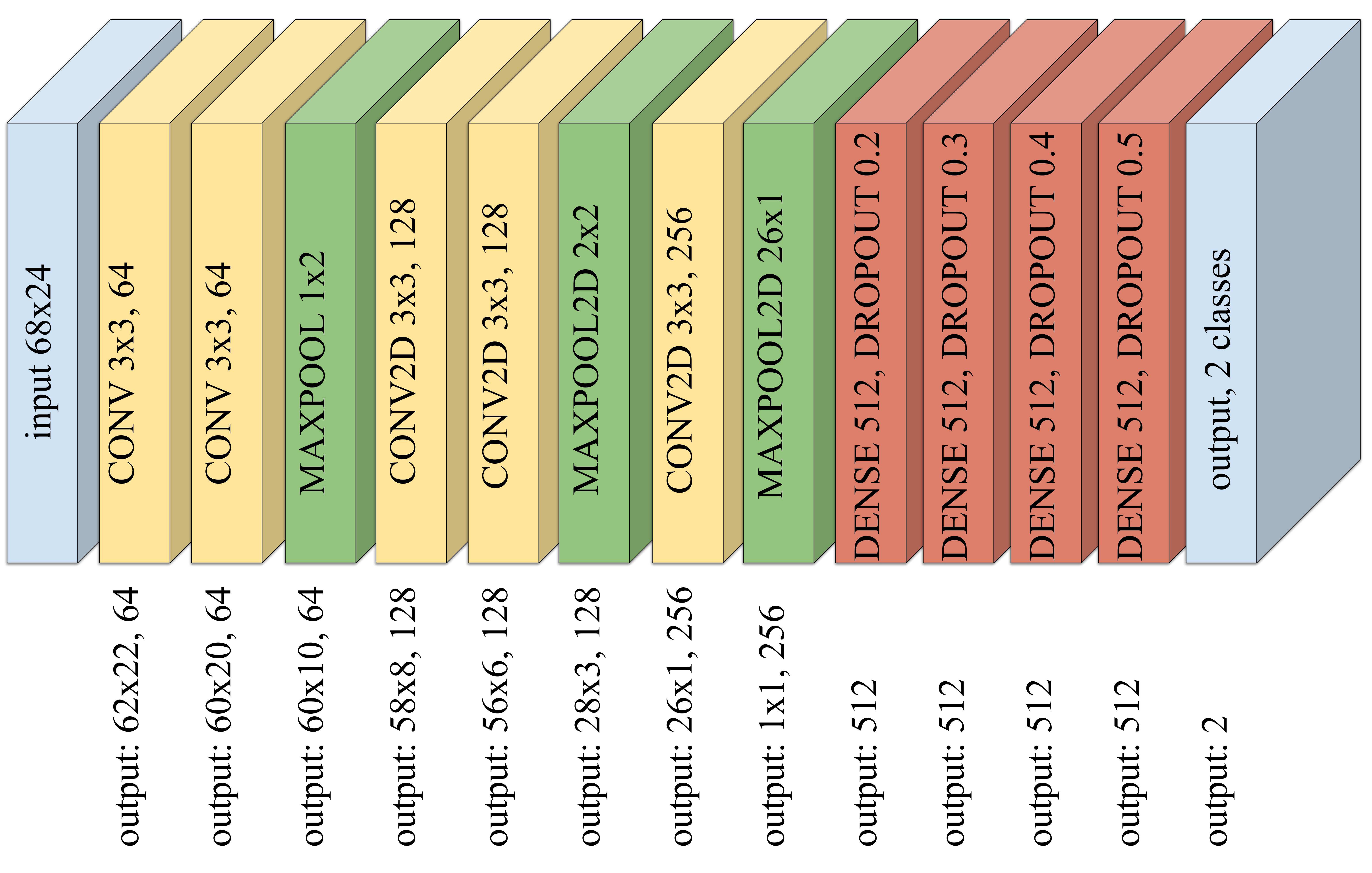}  
  \caption{Gender detection architecture}
  \label{fig:inaGender}
\end{subfigure}
\caption{(\subref{fig:inaMusic}) The depicted architecture addresses the speech/non-speech segment detection problem \cite{Doukhan:2018} (\subref{fig:inaGender}) The depicted architecture addresses gender detection of speech segments \cite{Doukhan:2018:2}. }
\label{fig:inaCNNs}
\end{figure*}

Since this study primarily aims to classify given segments into two classes, namely customer and agent, we do not perform a thorough speaker diarization, which requires a  neural network well-trained explicitly for this purpose. We instead use publicly available pre-trained neural networks for classifying given audio segments into four classes, i.e., speech, music, silence, and noise. It is also possible to identify the gender of speech segments. Both problems are addressed and solved using  CNNs in the joint studies \cite{Doukhan:2018} and \cite{Doukhan:2018:2}. inaSpeechSegmenter is an open-source implementation of these works where one can use these pre-trained CNNs. We use these networks to preprocess our audio recordings and create a training data set for the proposed purpose.  These CNNs are depicted in Figure \ref{fig:inaCNNs}. 

Prepocessing the call center recordings with the CNNs in Figure \ref{fig:inaCNNs} enables us to extract speech segments and their gender information accordingly. 
 
\section{Method}\label{sec:method}

\subsection{Data Set}\label{subsec:preparingdata}

Collecting and labeling data for deep supervised training is the most challenging and time-consuming step, and many studies in artificial intelligence literature are mostly about proposing new architectures for solving specific problems using publicly shared benchmark data sets like ImageNet \cite{Deng:2009}. 

The data used in this study is the call center recordings of the company. There are hundreds of thousands of recordings of varying lengths. For these recordings to be used as training data for our cause, we first filtered out very long and very short conversations. We only considered recordings of length between 60 and 600 seconds. Hence we obtained 48,462 files.

\begin{figure*}[]
    \centering
    \includegraphics[width=0.65\linewidth]{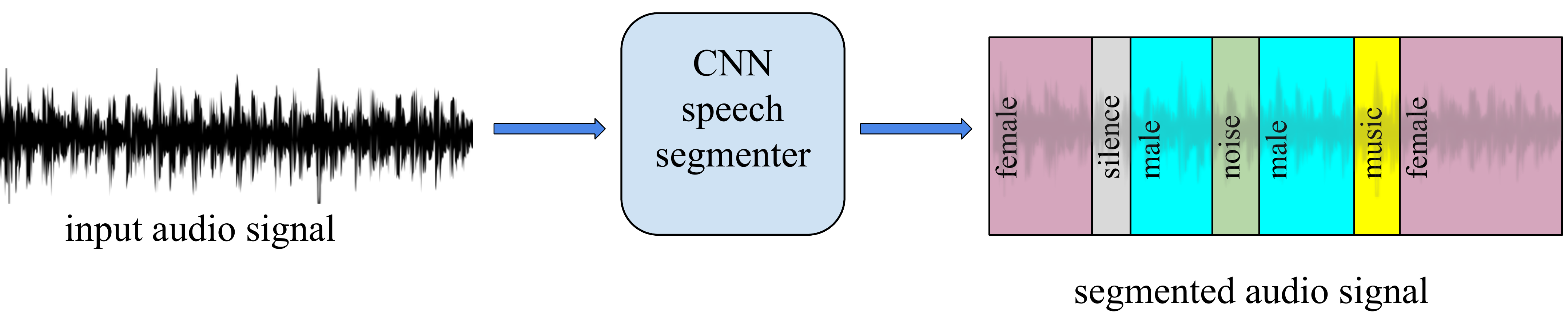}
    \caption{
    The first preprocessing step is to segment the single-channel audio signal into four groups representing ``silence", ``noise", ``speech" (with gender info), and ``music". inaSpeechSegmenter Tool \cite{Doukhan:2018},\cite{Doukhan:2018:2} is used to perform this segmentation.
    }
    \label{fig:input2segmented}
\end{figure*}

The inaSpeechSegmenter library can detect speech and non-speech segments also detect the gender of speech segments for a given audio sequence as depicted in Figure \ref{fig:input2segmented}. 

\begin{figure}[]
    \centering
    \includegraphics[width=0.99\linewidth]{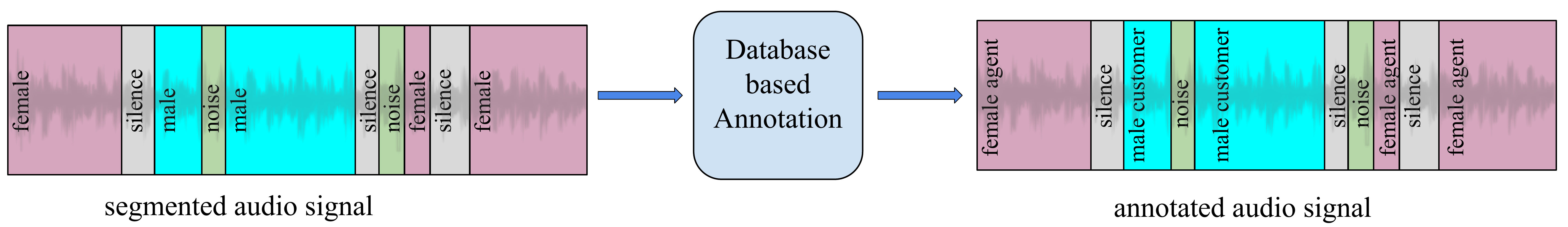}\\  
    \includegraphics[width=0.99\linewidth]{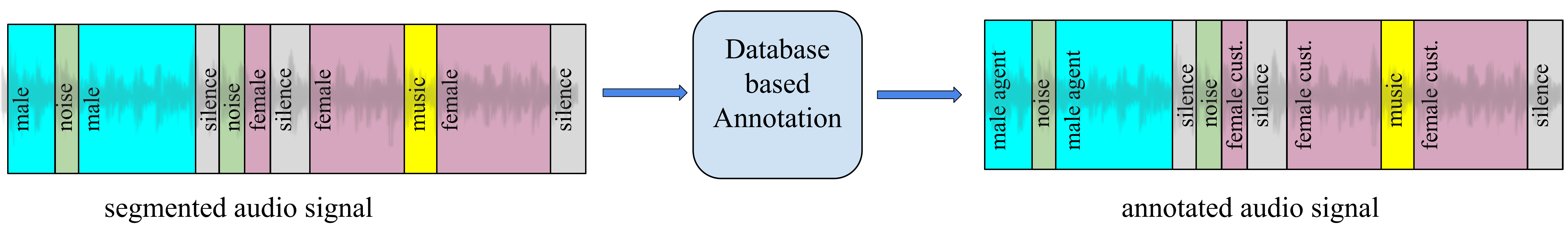}
    \caption{
Our Database based annotation scheme (DBAS) is depicted. Since we consider only the calls with two opposite gender segments, we mainly have two types of calls with the female agent - male customer, and the male agent - female customer pairs. 
    }
    \label{fig:segmented2annotated}
\end{figure}

The audio records where there exist two opposite genders are of our concern. Because we already have the agent id and agent gender information of any recording in our database and we can label speech segments with four different labels: ``female customer'', ``male customer'', ``female agent'', and ``male agent''. We call this phase ``Database based Annotation''. The database based annotation scheme (DBAS) with two possible cases, female agent - male customer, and male agent - female customer, is depicted in Figure \ref{fig:segmented2annotated}.

This annotation scheme theoretically finalizes our data picking procedure. On the contrary, we came up with a significant amount of ambiguous gender information for numerous speakers. inaSpeechSegmenter library randomly mislabels some male speakers' genders with aurally perceptible dominant high frequencies and some female speakers with dominant low frequencies in their voices. Hence we also filtered out the speakers where they are tagged with multiple gender labels in different conversations. We considered a speaker's conversations and discarded that speaker if she/he is labeled with multiple gender labels in those conversations, and kept the speakers who were consistently labeled with a single consistent gender label. 

Finally, the segments in conversations with two opposite genders can be labeled with female/male and customer/agent labels once the agent gender is known. We obtained 23,308 utterances from 377 different speakers, which total up to approximately 65 hours. Since we want the proposed network to learn speaker type, not the speaker voices, we spared 38 speakers for validation and did not use their utterances in the training epochs. These statistics are summarized in Table \ref{tab:acc}.

\begin{table*}[]
\caption{Table showing the number of speakers, utterances of different classes for training and validation data sets.} 
\centering 
\begin{tabular}{l c c l c c } 
\hline\hline 
 Data Type & \#Speakers &\#Utterances & Class & \#Speakers &\#Utterances \\ [0.5ex]
\hline 
&&&\\[-1ex]
 \raisebox{-1ex}{Train}  & \raisebox{-1ex}{339}   & \raisebox{-1ex}{20296}   &customer & 288 & 7969 \\
                                            &    &   & agent  & 51   & 12327\\
&&&&&\\[-1ex]
\hline
&&&&&\\[-1ex]
\raisebox{-1ex}{Validation}  & \raisebox{-1ex}{38}     & \raisebox{-1ex}{3012} & customer  & 32 & 801 \\
                                              &  &   & agent  & 6  & 2211 \\
&&&\\[-1ex]
\hline
\end{tabular}
\label{tab:acc}
\end{table*}

We kept each utterance length fixed at 10 seconds. All the records are originally kept in .gsm format with an 8kHz sampling frequency. We used log-amplitude mel-spectrograms as input since they have proven to outperform STFT,  MFCCs, and linear-amplitude mel-spectrograms \cite{Choi:2016}, \cite{Dieleman:2014}. As proposed in \cite{Choi:2017}, we used  96 mel-bins. The window and hop lengths are 200 and 80  samples, respectively.  With the above configuration, each utterance is transformed into a log-amplitude mel-spectrogram array with a size (96,1000).

\subsection{Proposed Model Architecture}\label{subsec:modelarch}

\begin{figure*}[]
    \centering
    \includegraphics[width=\linewidth]{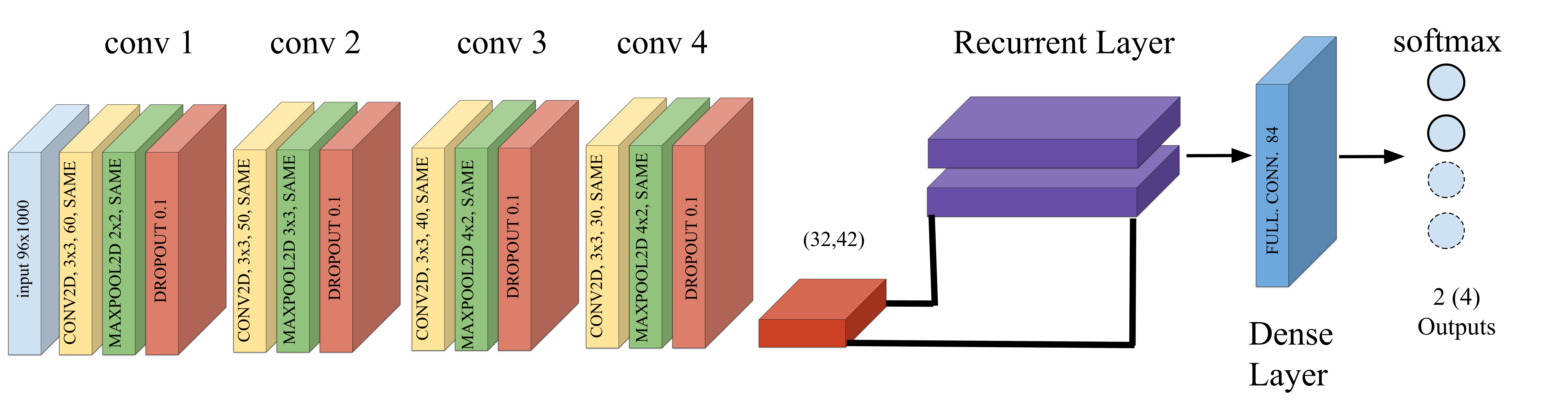}
    \caption{Inherited CRNN architecture for segment classification proposed in  \cite{Choi:2017},  and originally designed for music genre classification is depicted. }
    \label{fig:crnn}
\end{figure*}

For this study, we use a slightly modified version of the CRNN model proposed in \cite{Choi:2017} (see Figure \ref{fig:crnn}). In the architecture, four similarly structured convolution layers are followed by two recurrent layers whose output is connected to a fully connected dense layer and a softmax layer for the final classification.

Each convolutional layer has a cascaded max-pooling and dropout layer. Convolutional layers perform ``same convolution", and they have varying numbers of filters. Except for the first convolutional layer, the max-pool layers have a kernel size of (3,3). The dropout layer probabilities are kept at 0.1 for all four convolutional layers. 

The convolutional layers' output is a 2D array with a size (32,42), where the second dimension represents the temporal dimension. This output is directly fed to a recurrent neural network with two layers. GRU has become the most common choice for structuring RNNs. However, we tested both GRU and LSTM in the recurrent layer and reported their performances.

The last layer output of the recurrent layer is connected to a dense layer, which is succeeded by a softmax layer for final classification. For our two-class classification, the softmax layer has two nodes to classify customer and agent segments. The architecture performs well on both training and validation data sets for this problem. We report the results in  Section \ref{sec:results}.

\subsection{Gender Extension}\label{subsec:genderext}

Our main goal is to classify audio segments into two classes, namely customer and agent. Hence, we prepared the training and validation data sets with these labels accordingly. However, we inherently have gender data either and we can re-label the data sets with four labels: ``female customer'', ``male customer'', ``female agent'', and ``male agent''. With these new labels, the data statistics in Table \ref{tab:acc} can be modified as Table \ref{tab:acgc}.

\begin{table*}[]
\caption{Table showing the number of speakers, utterances of different genders, and classes for training and validation data sets.} 
\centering 
\begin{tabular}{l c l c l c} 
\hline\hline 
 Data Type & \#Speak.-\#Utter. & Class & \#Speak.-\#Utter. & Gender &  \#Speak.-\#Utter. \\ [0.5ex]
\hline 
&&&\\[-1ex]
& & & & female & 124 - 4434\\
 \raisebox{-1ex}{Train}  & \raisebox{-1ex}{339-20296}  & \raisebox{1ex}{customer}  & \raisebox{1ex}{288-7969} & male & 164-3535 \\
&   & \raisebox{-1ex}{agent}  & \raisebox{-1ex}{51-12327} & female & 30-6974\\
& & & & male & 21-5353\\                                     &&&\\[-1ex]
\hline
&&&\\[-1ex]
& &  & & female & 12-264\\
 \raisebox{-1ex}{Test}  & \raisebox{-1ex}{38-3012}   & \raisebox{1ex}{customer}  & \raisebox{1ex}{32-801} & male & 20-537 \\
&   & \raisebox{-1ex}{agent}     & \raisebox{-1ex}{6-2211} &female & 3-1870\\
& & & & male & 3-341\\                                            
&&&\\[-1ex]
\hline
\end{tabular}
\label{tab:acgc}
\end{table*}

We trained our network with four labels and observed that the model also performs well on the training and validation datasets. The only modification in the model is at the softmax layer, where we increase the number of output nodes from two to four. The corresponding training performances are reported in  Section \ref{sec:results}.

\subsection{Implementation Details}

We implemented our codes in Python  \cite{Rossum:1995}. While structuring our CRRN, we used Keras \cite{Chollet:2015} interface for Tensorflow 2.1 \cite{TensorFlow:2015} library. Other common tools such as Pandas \cite{Pandas} and Numpy \cite{Numpy:2008}are also used for managing data, Matplotlib \cite{Matplotlib:2007} and Scikit \cite{Scikit:2011} libraries are used for data visualization. 

The experiments are performed on a Dell Precision 5820 workstation with a 16 core 3.70GHz  Intel(R) Xeon(R) W-2145 CPU, 64GB 2666MHz DDR4 RAM, and a GeForce RTX 2080 8GB GPU.  Since it is not possible to load and process the whole data, we used the data pipeline procedure of Tensorflow.  For convenience, we kept every training sample in this folder structure: {\tt <train/validation>/<agent/customer>/ <female/male>/<speakerId>/<utteranceNo>.npy} 

Tensorflow's data pipelines make it possible to read each data file and extract labels from the folder names at processing time. We used Adam optimizer \cite{Kingma:2014} and early stopping criteria \cite{Yao:2007} for the optimization. 

\section{Results}\label{sec:results}

We dedicate this section to the experimental results. Recalling Figure \ref{fig:crnn}, we construct our CRNN first using GRUs in the recurrent layer and tested it on two problems, presented in Section \ref{sec:method}. We then replaced GRUs with LSTMs and similarly tested the network with the same data and problems. We call our original problem as ``2 Class Case'' and the gender extended version of the problem defined in Section \ref{subsec:genderext} to as ``4 Class Case''.

\begin{figure*}[]
\begin{subfigure}{.50\textwidth}
  \centering
  \includegraphics[width=\linewidth]{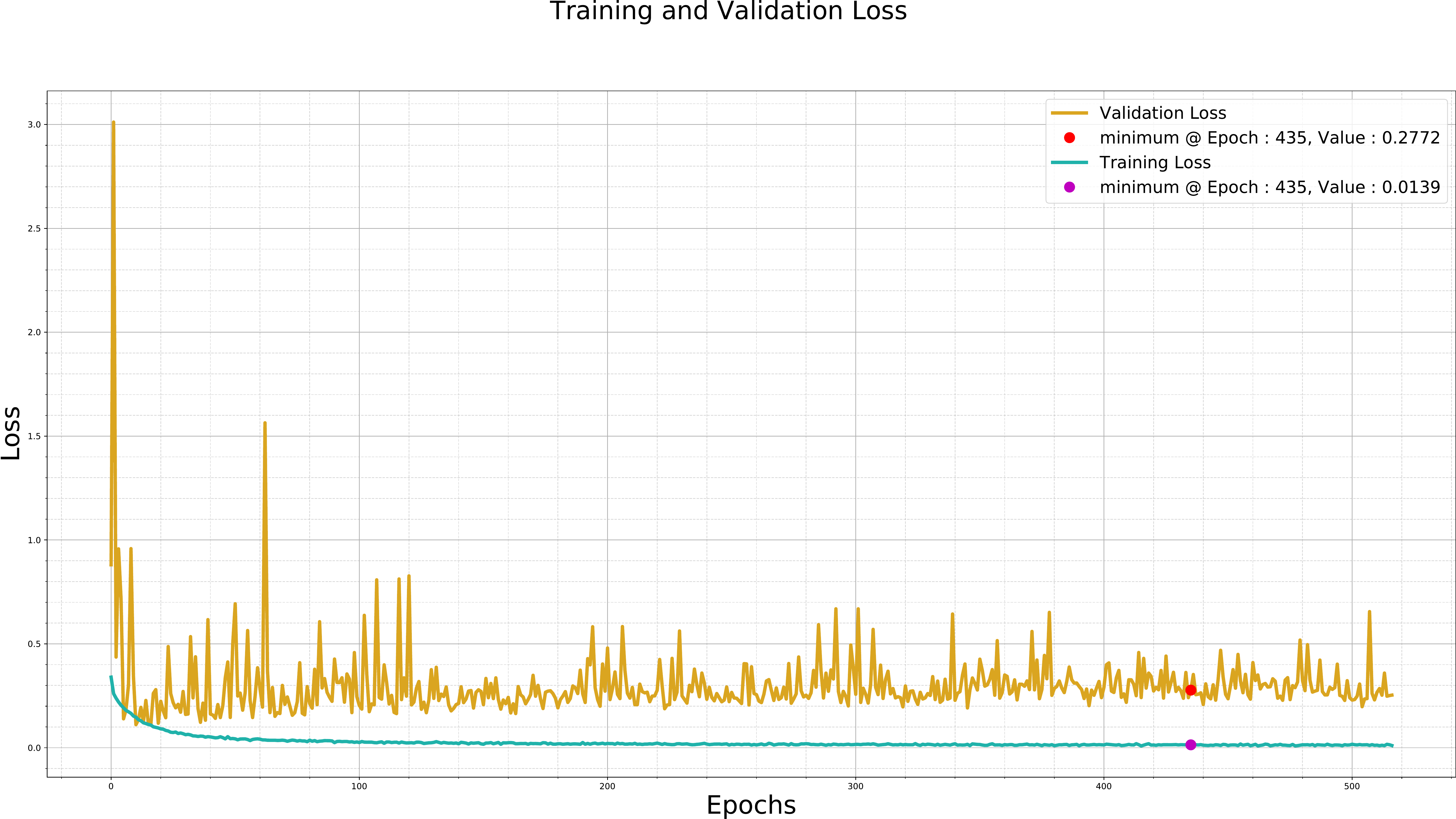}  
  \caption{ Loss of GRU with 2 Class Case}
  \label{fig:gru2closs}
\end{subfigure}
\begin{subfigure}{.50\textwidth}
  \centering
  \includegraphics[width=\linewidth]{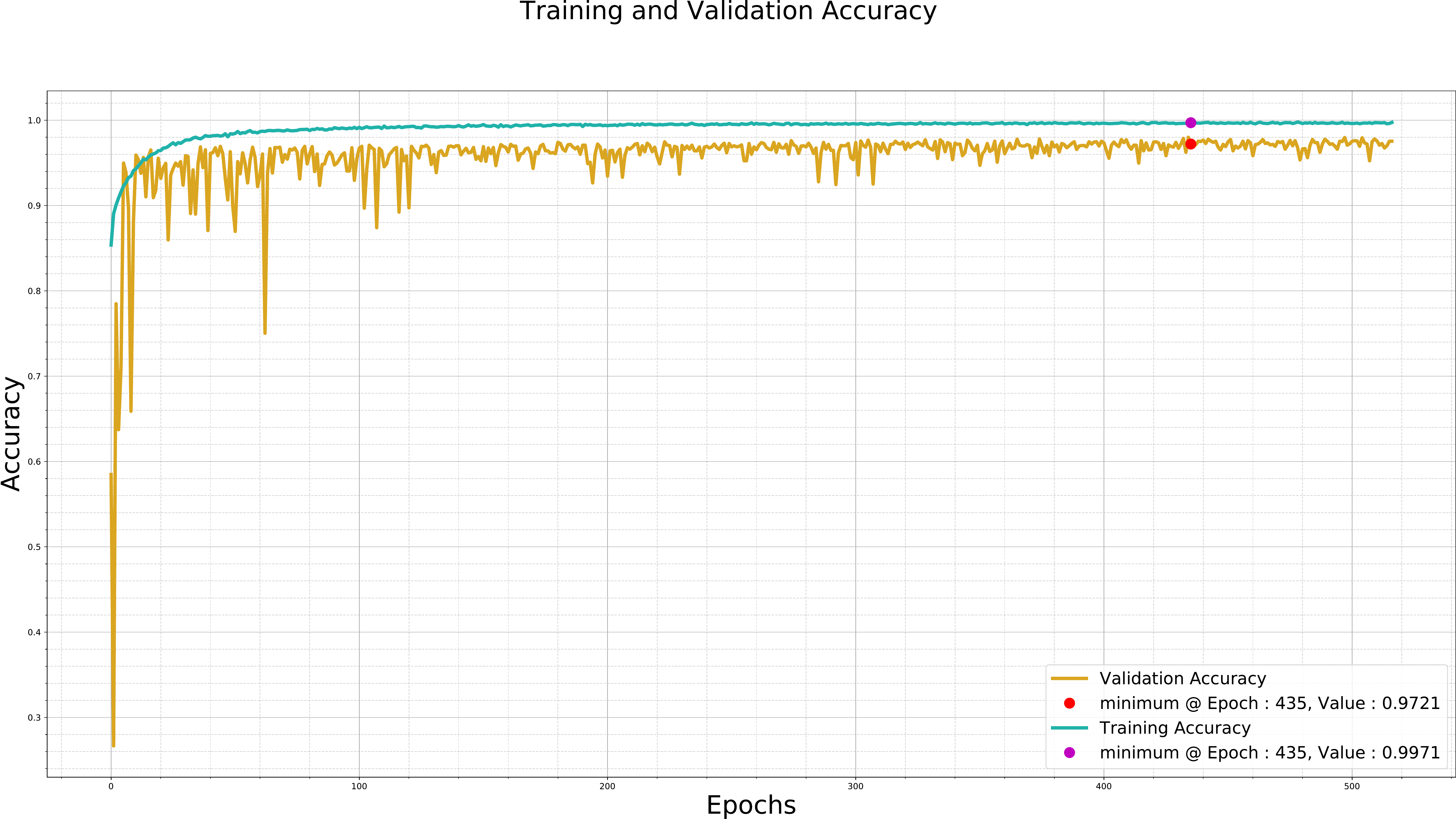}  
  \caption{Accuracy of GRU with 2 Class Case}
  \label{fig:gru2cacc}
\end{subfigure}\\
\begin{subfigure}{.50\textwidth}
  \centering
  \includegraphics[width=\linewidth]{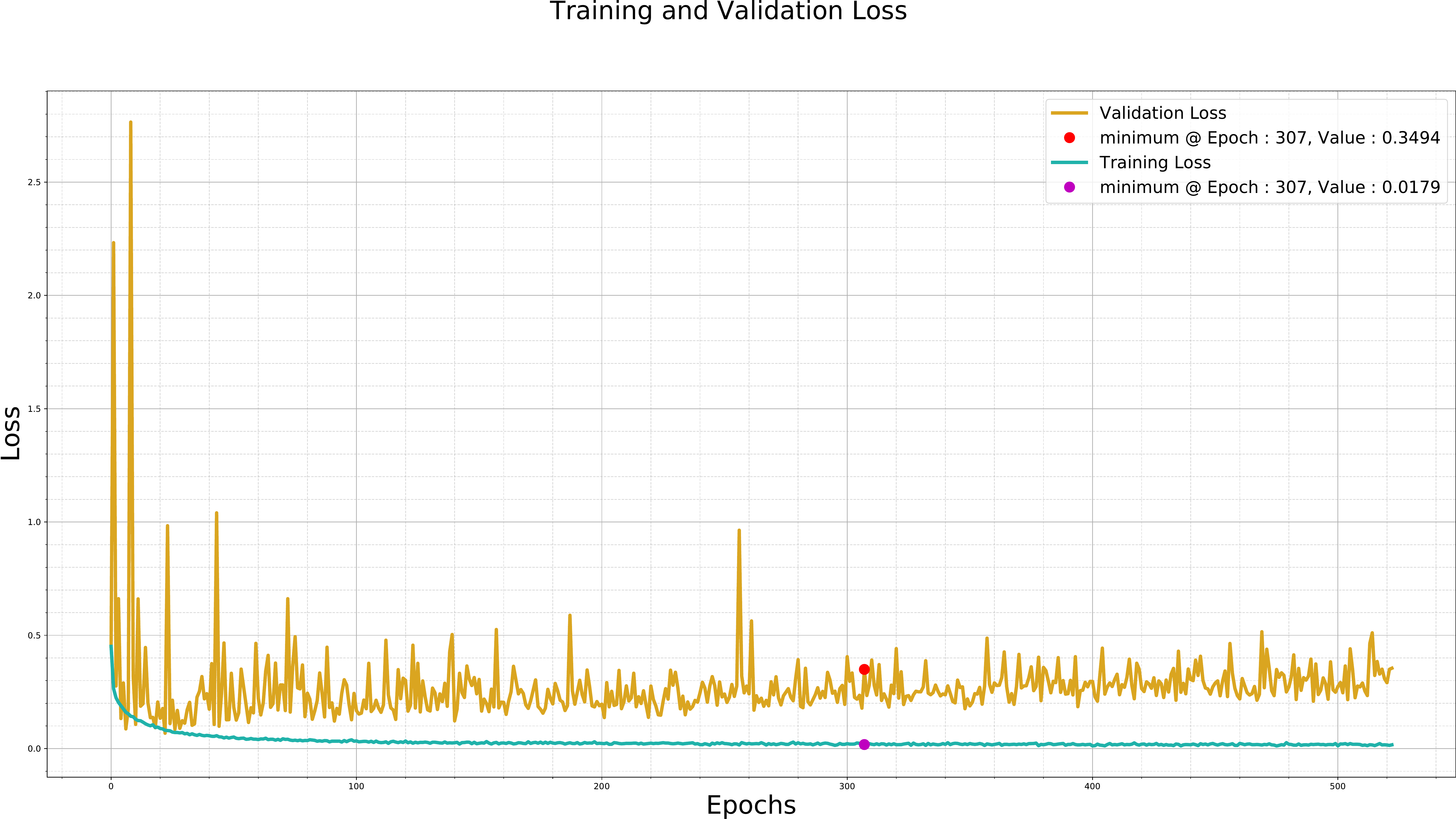}  
  \caption{ Loss of GRU with 4 Class Case  }
  \label{fig:gru4closs}
\end{subfigure}
\begin{subfigure}{.50\textwidth}
  \centering
  \includegraphics[width=\linewidth]{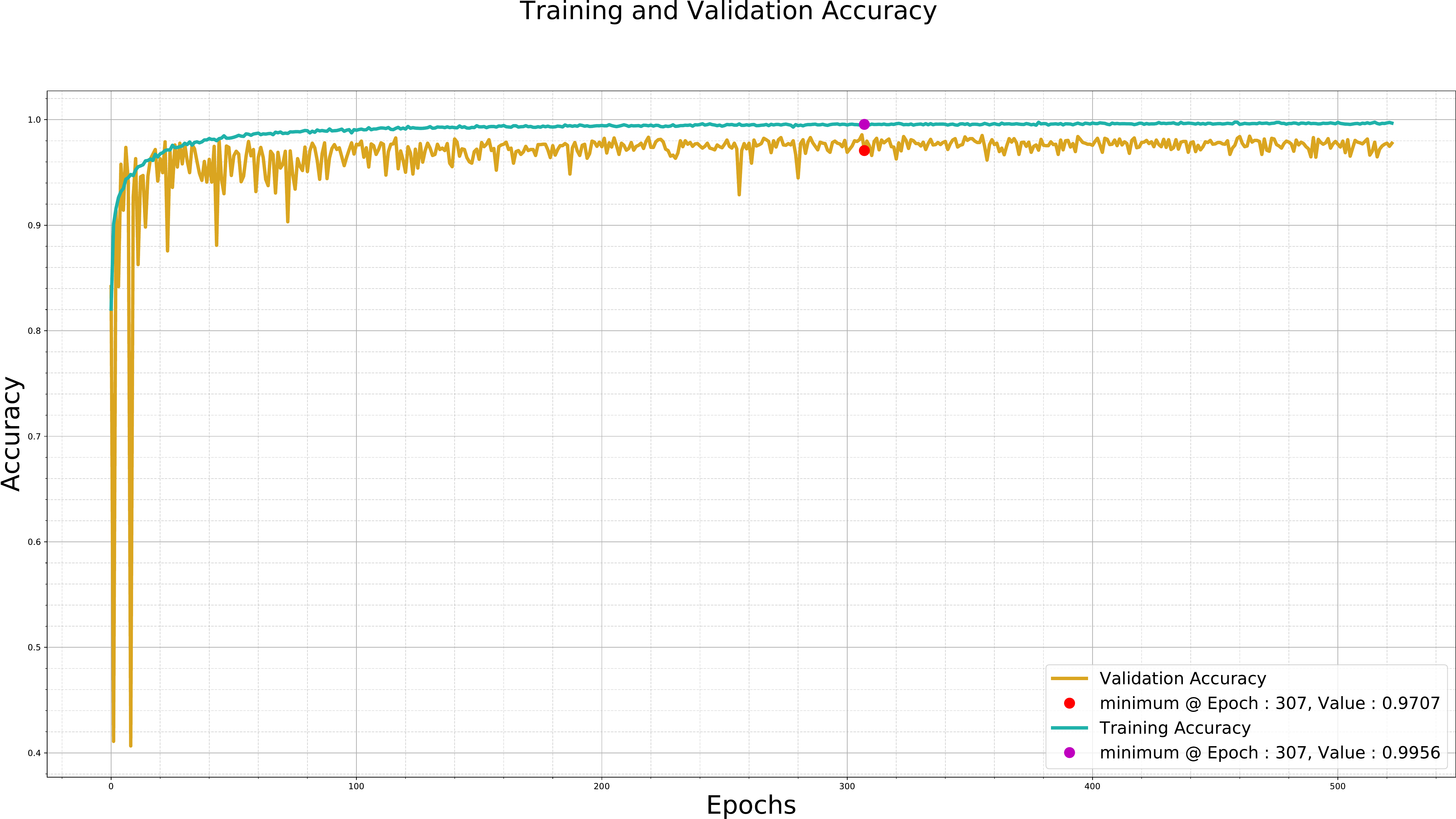}  
  \caption{Accuracy of GRU with 4 Class Case}
  \label{fig:gru4cacc}
\end{subfigure}\\
\caption{Training performance plots of the network in Figure \ref{fig:crnn} with the recurrent layer built with GRUs. (\subref{fig:gru2closs}), (\subref{fig:gru2cacc}) Loss and accuracy plots of training epochs for 2 Class Case, respectively. (\subref{fig:gru4closs}),(\subref{fig:gru4cacc}) Loss and accuracy plots of training epochs for 4 Class Case, respectively. Epochs at which the best accuracies are achieved have been indicated in each plot. }
\label{fig:gruresults}
\end{figure*}

\begin{figure*}[]
\begin{subfigure}{.50\textwidth}
  \centering
  \includegraphics[width=\linewidth]{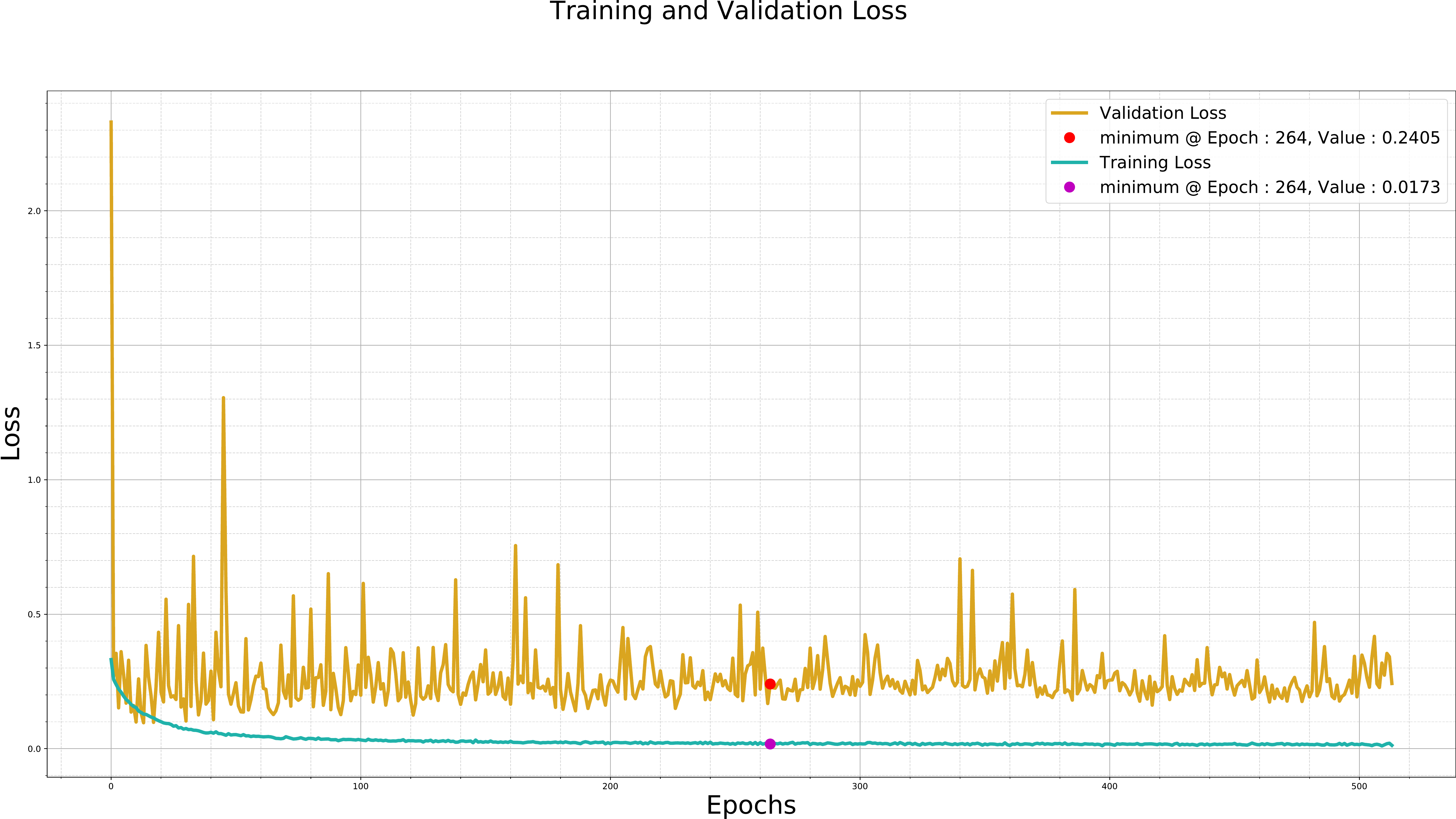}  
  \caption{Loss of LSTM with  2 Class Case}
  \label{fig:lstm2closs}
\end{subfigure}
\begin{subfigure}{.50\textwidth}
  \centering
  \includegraphics[width=\linewidth]{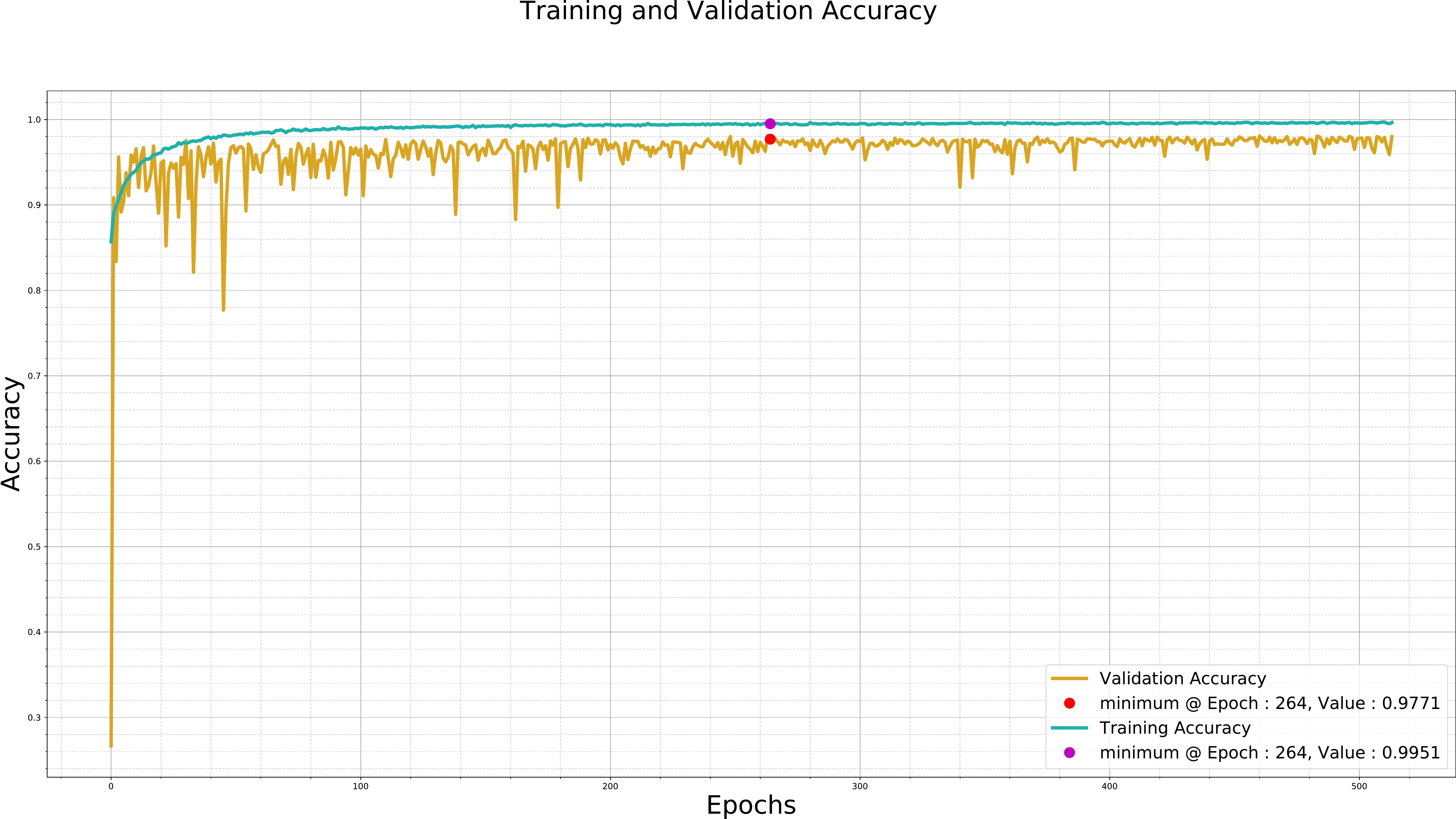}  
  \caption{Accuracy of LSTM with 2 Class Case }
  \label{fig:lstm2cacc}
\end{subfigure}\\
\begin{subfigure}{.50\textwidth}
  \centering
  \includegraphics[width=\linewidth]{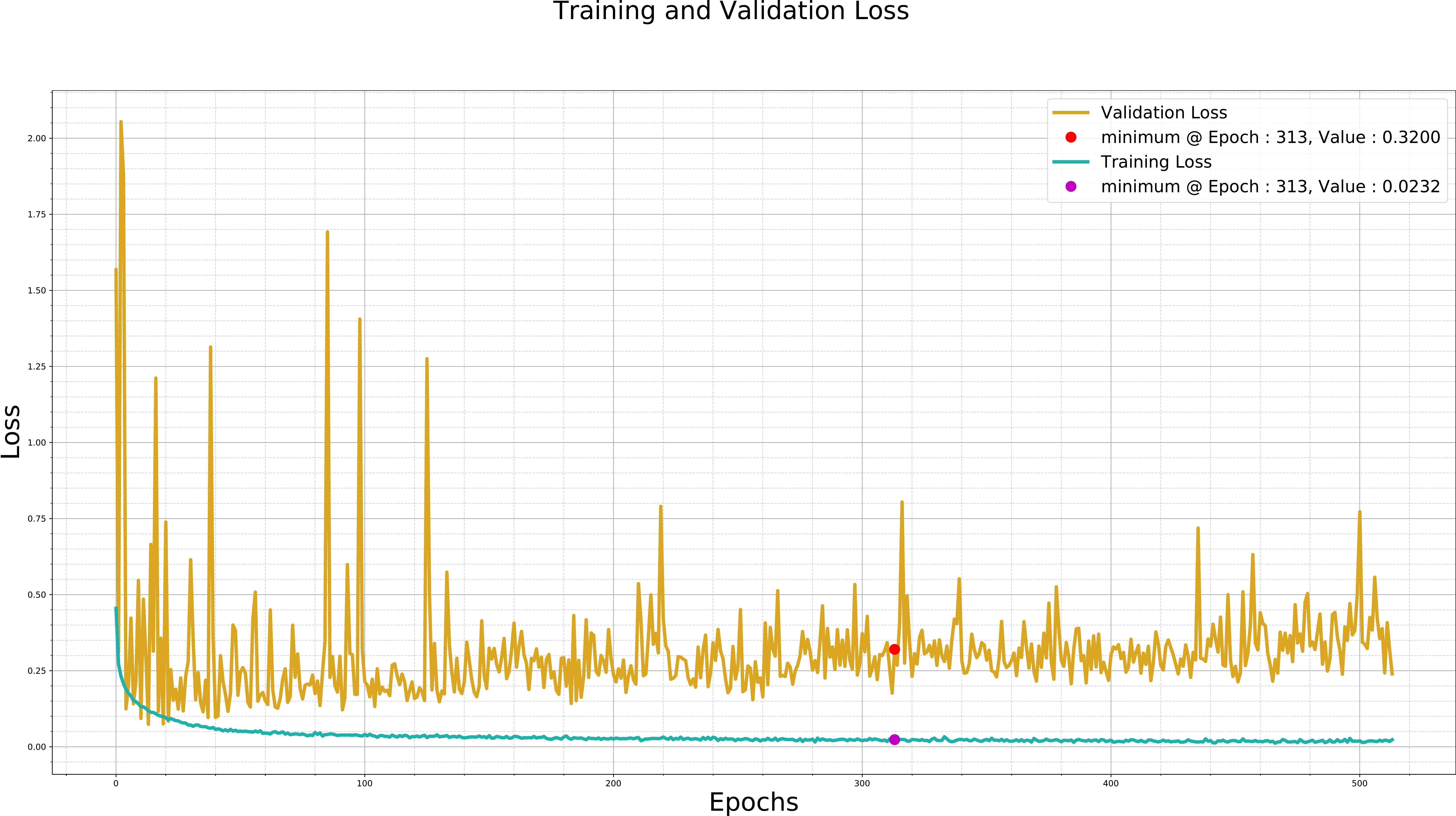}  
  \caption{ Loss of LSTM with 4 Class Case }
  \label{fig:lstm4closs}
\end{subfigure}
\begin{subfigure}{.50\textwidth}
  \centering
  \includegraphics[width=\linewidth]{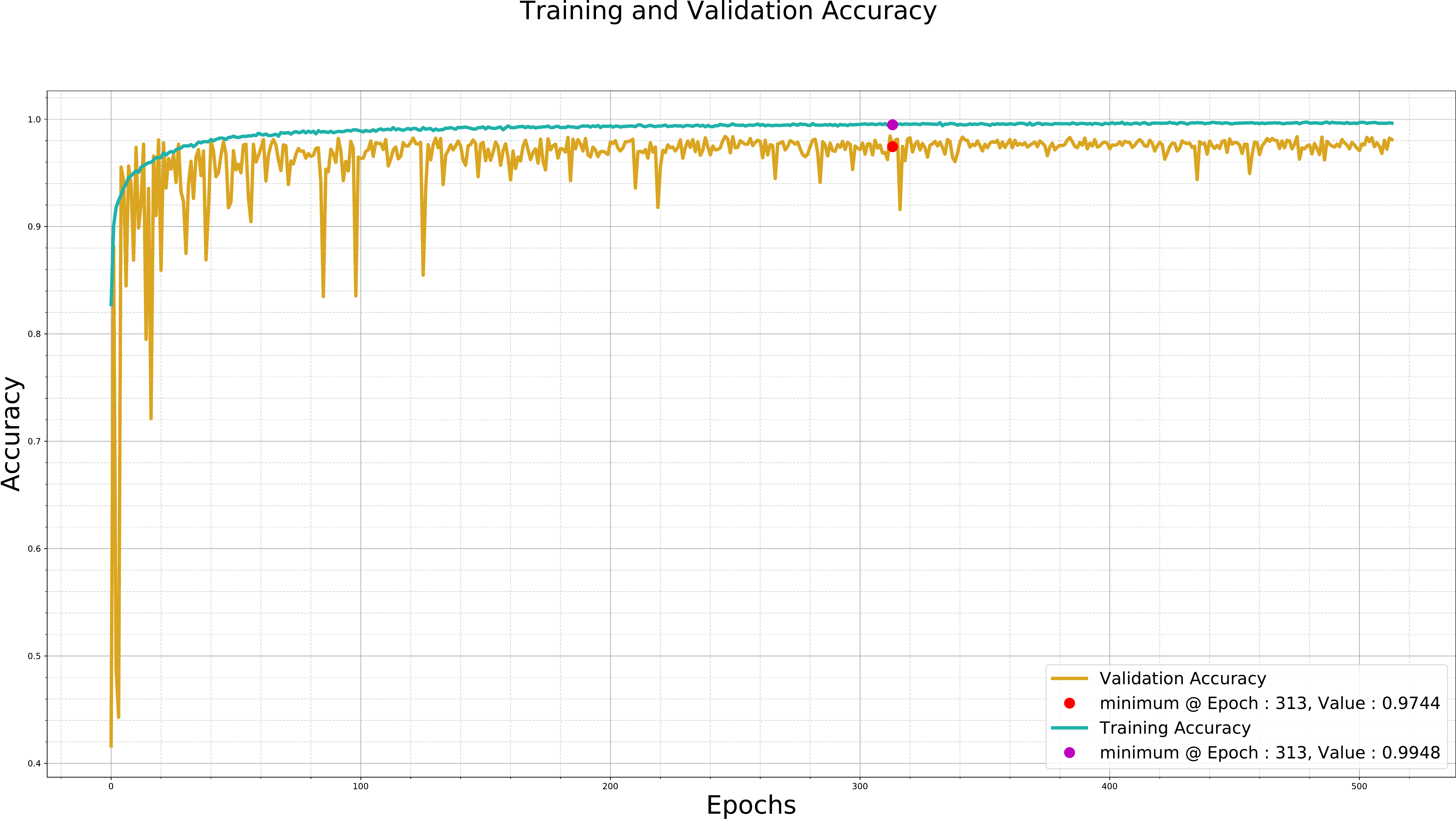}  
  \caption{Accuracy of LSTM with 4 Class Case }
  \label{fig:lstm4cacc}
\end{subfigure}
\caption{Training performance plots of the network in Figure \ref{fig:crnn} with the recurrent layer built with LSTMs. (\subref{fig:lstm2closs}), (\subref{fig:lstm2cacc}) Loss and accuracy plots of training epochs for 2 Class Case, respectively. (\subref{fig:lstm4closs}),(\subref{fig:lstm4cacc}) Loss and accuracy plots of training epochs for 4 Class Case, respectively.. Epochs at which the best accuracies are achieved have been indicated in each plot.}
\label{fig:lstmresults}
\end{figure*}

In Figure \ref{fig:gruresults}, we list experiments' results using GRUs in the recurrent layer.  Figures \ref{fig:gru2closs}, \ref{fig:gru2cacc},   \ref{fig:gru4closs}, and \ref{fig:gru4cacc} present the network performance for problems. Similarly, Figures \ref{fig:lstm2closs}, \ref{fig:lstm2cacc},   \ref{fig:lstm4closs}, and \ref{fig:lstm4cacc} present the same results for the network with LSTM units in the recurrent layer. 

We see that both network architectures perform quite well on both training and validation data sets for either problem. However, it is essential to see the network's classification performance for each class separately. For this reason, we plotted confusion matrices for the experimental results in Figure 7. The labeling conventions for the experiments can be summarized in  Table \ref{tab:labelconv}.

\begin{table}[]
\caption{Labeling Convention} 
\centering 
\begin{tabular}{l c c c c c c} 
\hline\hline 
Problem & Label Number &Label Name \\ [0.5ex]
\hline 
2 Class &0 & customer \\
 &1 & agent\\
\hline
4 Class &0 & female customer \\
&1 & male customer\\
&2 & female agent\\
&3 & male agent\\
\hline
\end{tabular}
\label{tab:labelconv}
\end{table}

\begin{figure*}[]
\begin{subfigure}{.24\textwidth}
  \centering
  \includegraphics[width=\linewidth]{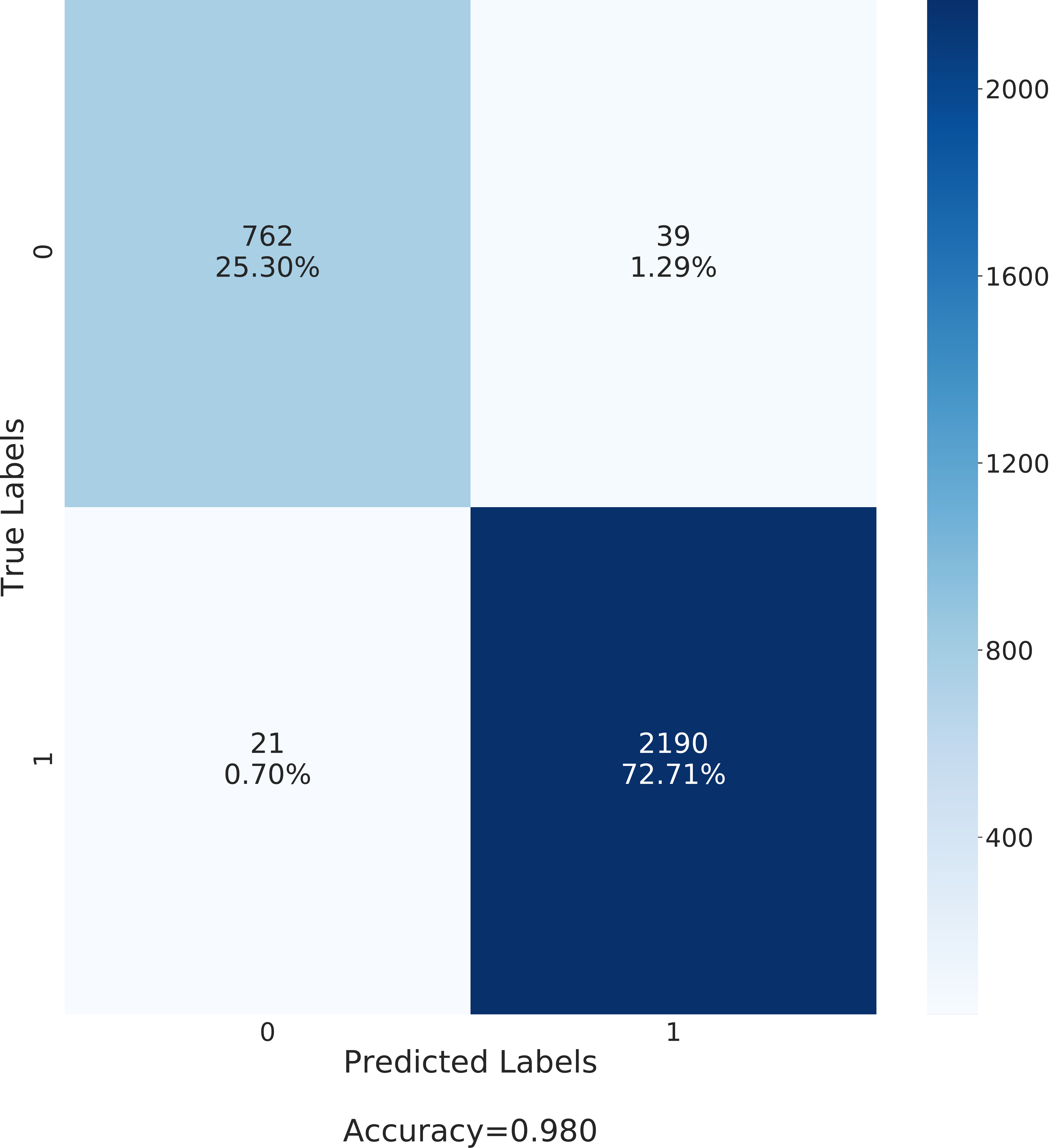}  
  \caption{}
  \label{fig:gru2cconf}
\end{subfigure}
\begin{subfigure}{.24\textwidth}
  \centering
  \includegraphics[width=\linewidth]{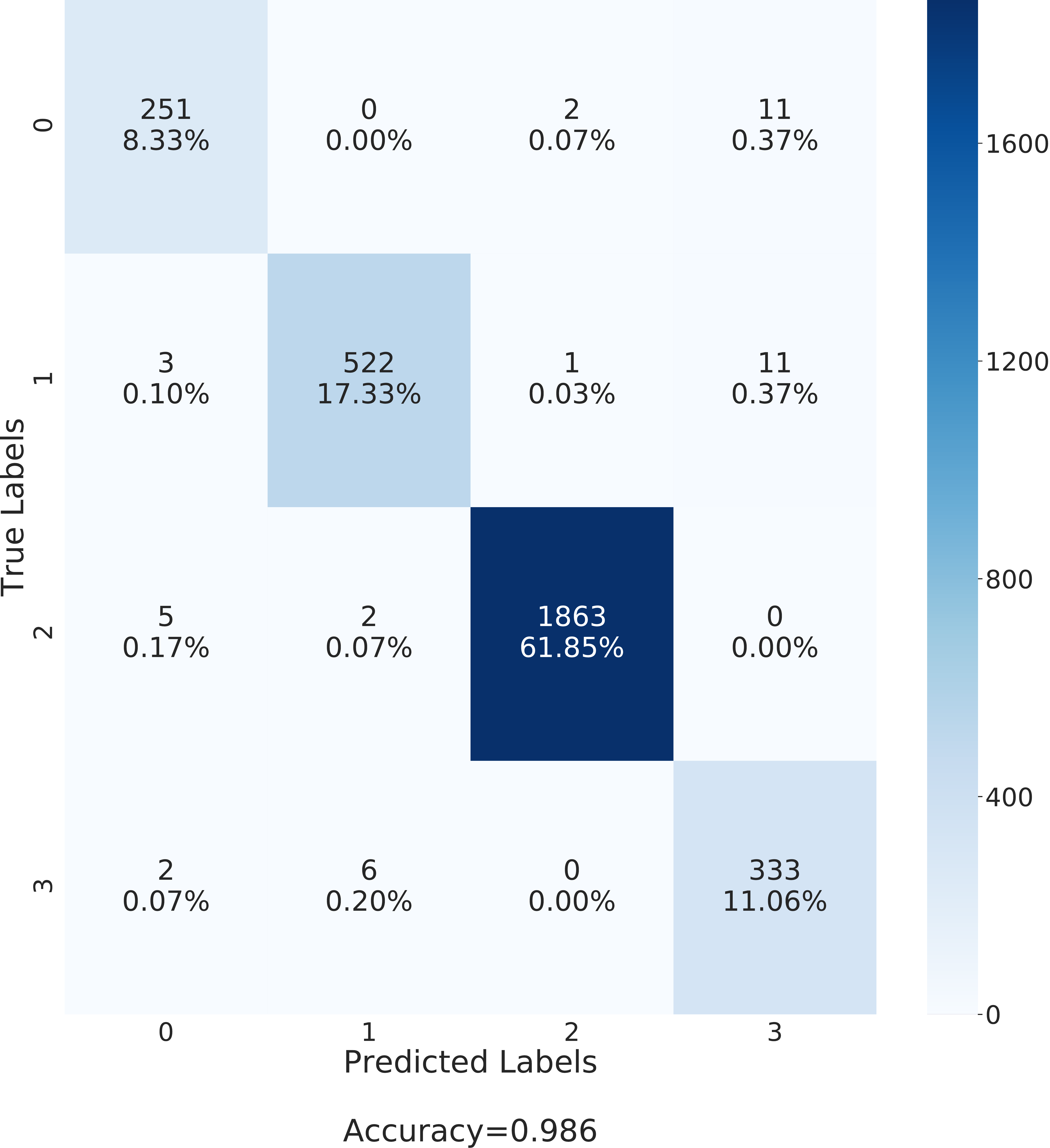}  
  \caption{}
  \label{fig:gru4cconf}
\end{subfigure}
\begin{subfigure}{.24\textwidth}
  \centering
  \includegraphics[width=\linewidth]{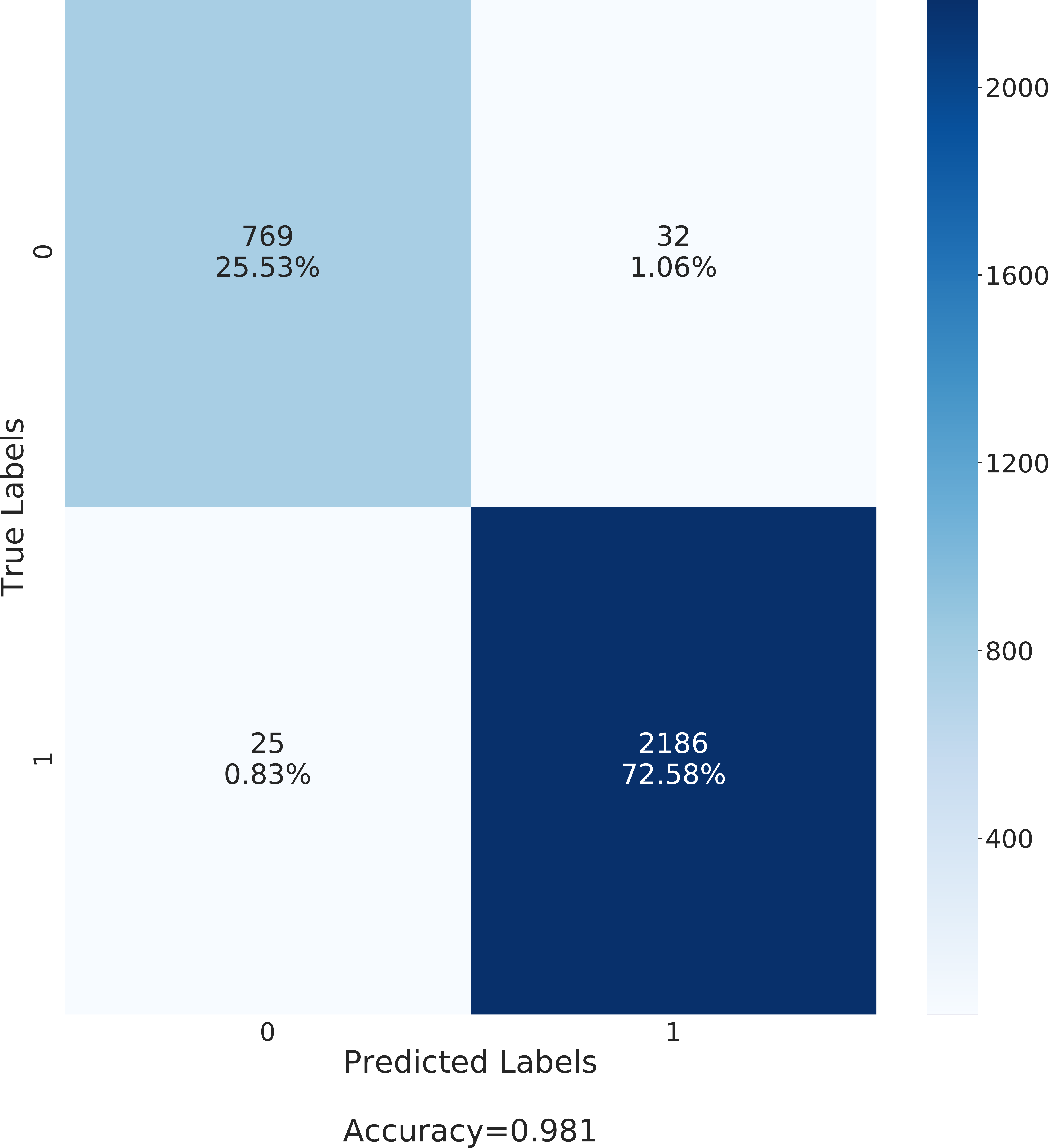}  
  \caption{}
  \label{fig:lstm2cconf}
\end{subfigure}
\begin{subfigure}{.24\textwidth}
  \centering
  \includegraphics[width=\linewidth]{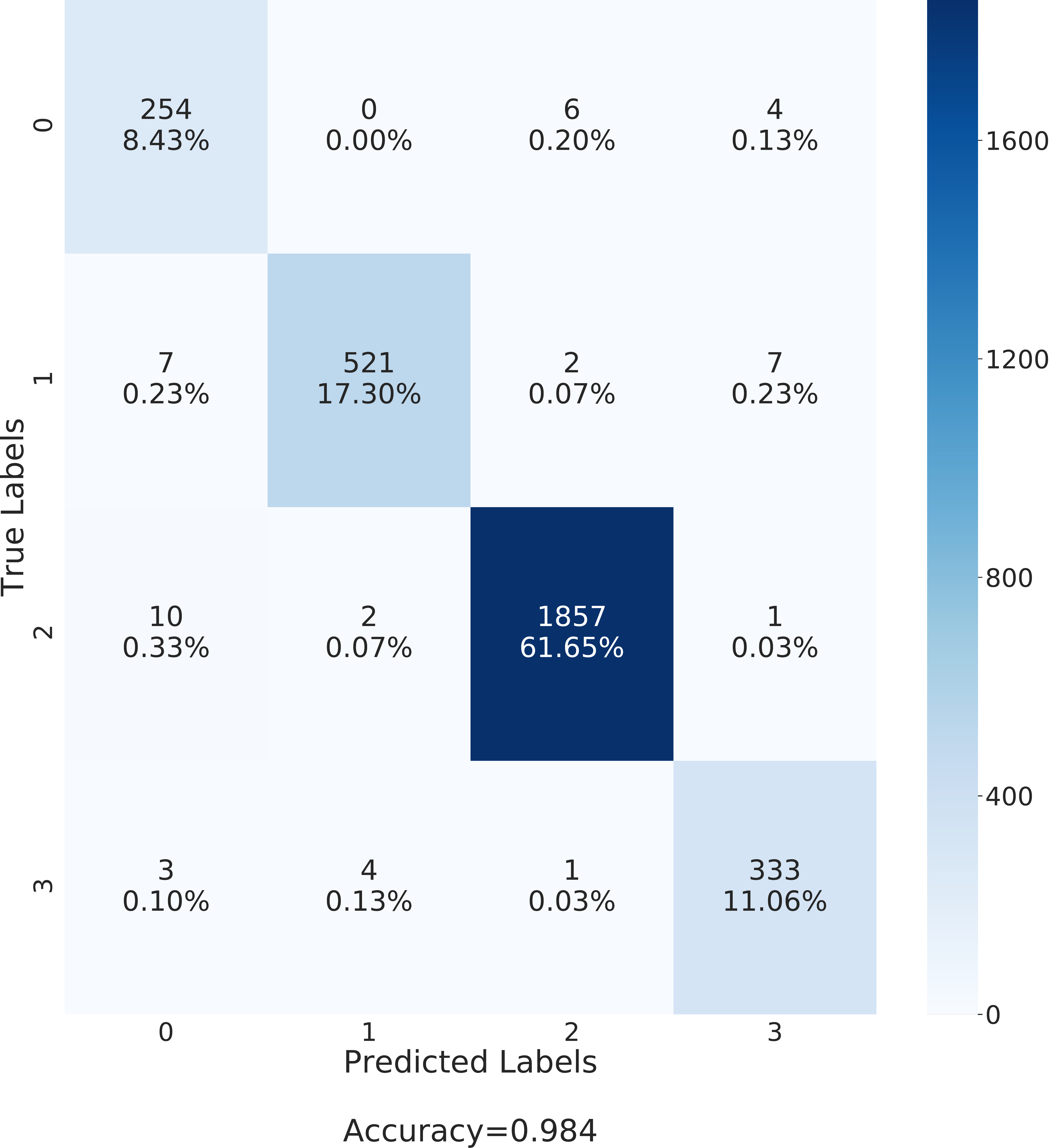}  
  \caption{}
  \label{fig:lstm4cconf}
\end{subfigure}\\
\caption{Confusion matrices of the experiments in Figures \ref{fig:gruresults} and \ref{fig:lstmresults}. (\subref{fig:gru2cconf}), (\subref{fig:gru4cconf}) Confusion matrices of the 2 Class and 4 Class Cases for the GRU network respectively. (\subref{fig:lstm2cconf}), (\subref{fig:lstm4cconf}) Confusion matrices of the 2 Class and 4 Class Cases for the LSTM network respectively.}
\label{fig:confmatrices}
\end{figure*}

The classification performance of the network can also be measured with a numerical metric called F1 Score. This score is calculated by using precision and recall rates. These sub-metrics are calculated by using true-positive (TP), false-positive (FP), and false-negative (FN) values, which can be deduced from the confusion matrices. Precision, Recall, and F1 Score can be defined as Equations \ref{eq:precision}, \ref{eq:recall}, and \ref{eq:f1}, respectively. F1 score can be calculated for each class separately.
 
\begin{align}
    \text{Precision} &= \frac{TP}{TP+FP}\label{eq:precision}\\
    \text{Recall} &= \frac{TP}{TP+FN}\label{eq:recall}\\
    \text{F1} &= 2*\frac{\text{Precision}*\text{Recall}}{\text{Precision}+\text{Recall}}\label{eq:f1}
\end{align}

We summarize the experimental results in Table  \ref{tab:grulstmcomp}. This table shows us that the GRU network's number of parameters is less than the LSTM network. By looking at the epoch number at which we achieve maximum accuracy, we see that the GRU network tends to be trained more before it overfits. By looking at the maximum accuracy values achieved for the validation data, we can say that both networks perform similarly.  We show F1 Scores calculated for each class for the corresponding problems. Even the training data set is not well-balanced, the F1 scores are high enough to deduce that the network achieves a performance close to human-level performance (HLP) for the given problems. 

\begin{table*}[]
\caption{GRU - LSTM Comparison} 
\centering 
\begin{tabular}{l c c c c l} 
\hline\hline 
 Unit Type & Problem &\#Params. & Max.Acc.Epoch & Train.- Valid. Acc. & F1 Scores \\ [0.5ex]
\hline 
 GRU &2 Class &118,478  &435  &0.9972 - 0.9721 &[0.9621, 0.9864] \\
 &4 Class & 118,648 &307 &0.9956 - 0.9702& [0.9561, 0.9784, 0.9973, 0.9568 ]\\
\hline
 LSTM &2 Class &135,782 &264 &0.9951 - 0.9771 &[0.9642, 0.9871]\\
 &4 Class &135,952 &313 &0.9948 - 0.9744 &[0.9442, 0.9793, 0.9941, 0.9708 ]\\
\hline
\end{tabular}
\label{tab:grulstmcomp}
\end{table*}

\subsection{Can the network generalize to other languages?} \label{subsec:generalization}

We showed that the proposed network architecture gives significant results on both training and validation data sets. The data set has thousands of different utterances from different speakers. Nevertheless,  all of the utterances belong to native Turkish speakers. On the other hand, we thought that the call center agents somehow have a universal  style of speaking regardless of their language. The same intuition can also be extended to the customers. Hence, we prepared a testbed for testing our network's generalization capability to other languages. 

\begin{figure*}[]
\begin{subfigure}{.50\textwidth}
  \centering
  \includegraphics[width=\linewidth]{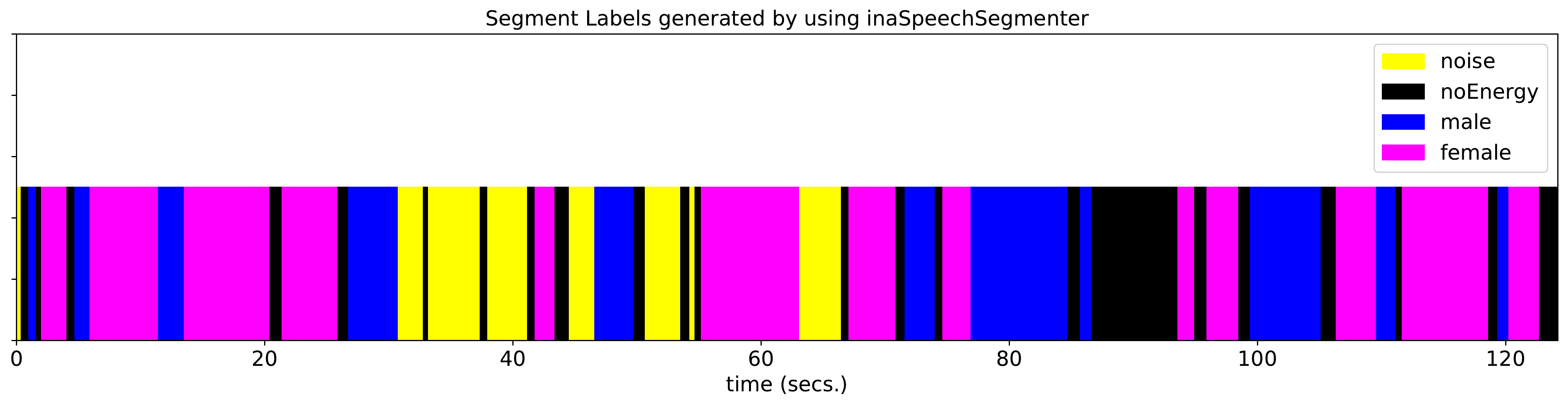}  
  \caption{ Segments created by inaSpeechSegmenter }
  \label{fig:inaLabelsGerman1}
\end{subfigure}
\begin{subfigure}{.50\textwidth}
  \centering
  \includegraphics[width=\linewidth]{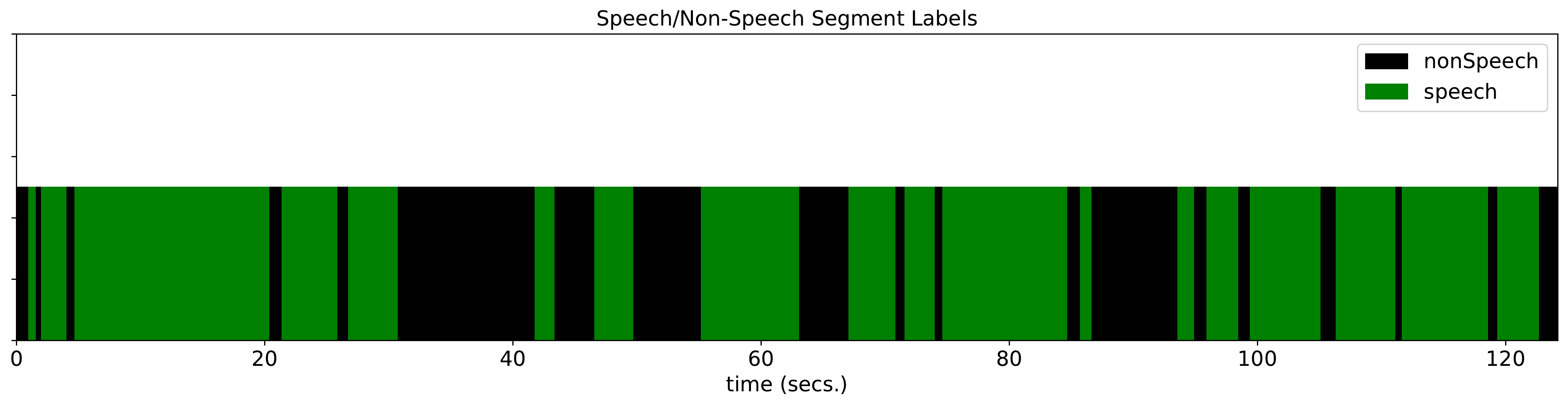}  
  \caption{Speech, Non-speech segments obtained using \ref{fig:inaLabelsGerman1} }
  \label{fig:speechLabelsGerman1}
\end{subfigure}\\
\begin{subfigure}{.50\textwidth}
  \centering
  \includegraphics[width=\linewidth]{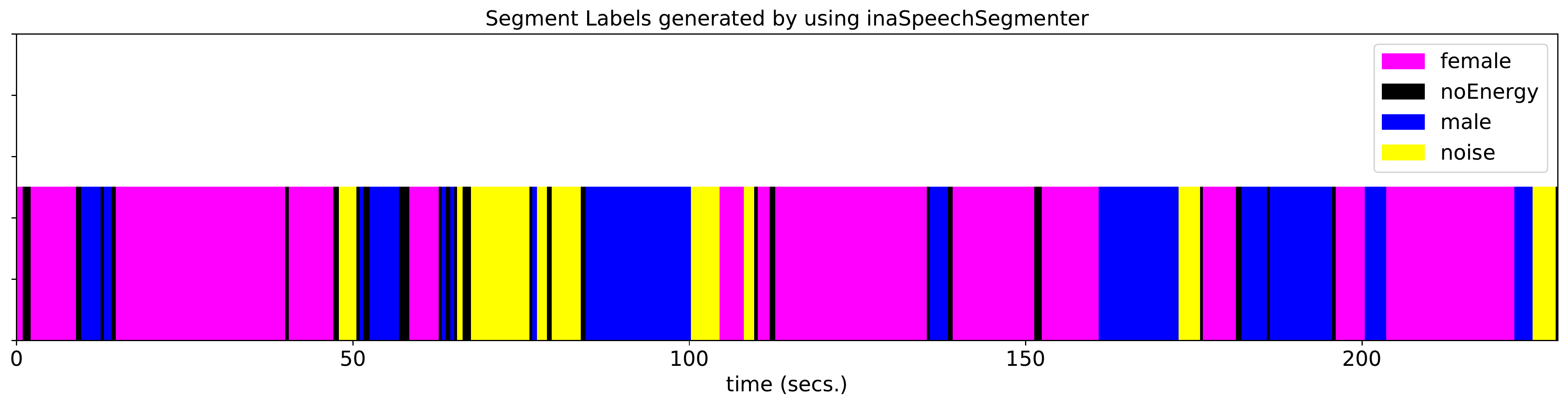}  
  \caption{ Segments created by inaSpeechSegmenter }
  \label{fig:inaLabelsGerman2}
\end{subfigure}
\begin{subfigure}{.50\textwidth}
  \centering
  \includegraphics[width=\linewidth]{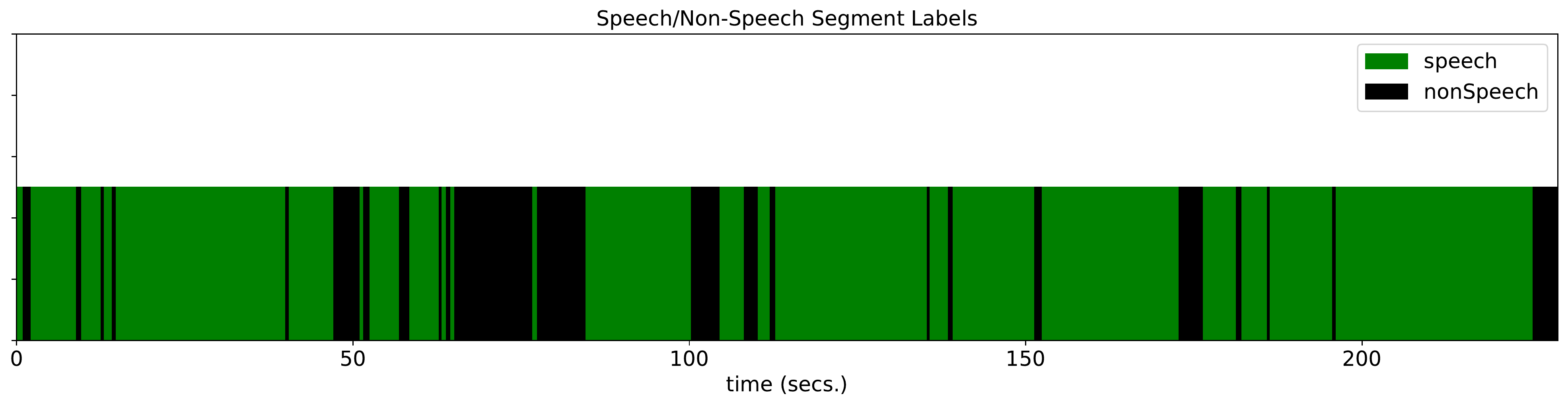}  
  \caption{Speech, Non-speech segments obtained using \ref{fig:inaLabelsGerman2}}
  \label{fig:speechLabelsGerman2}
\end{subfigure}\\
\caption{ Segments detected using German test samples. (\subref{fig:inaLabelsGerman1}), (\subref{fig:inaLabelsGerman2}) Shows the segment labels detected using inaSpeechSegmenter. (\subref{fig:speechLabelsGerman1}), (\subref{fig:speechLabelsGerman2}) Speech - Non-speech segments extracted by using (\subref{fig:inaLabelsGerman1}), (\subref{fig:inaLabelsGerman2}), respectively.
}
\label{fig:germanSamples}
\end{figure*}

In the first test, we have chosen two sales call between one of the company's native German speaker female agent and two native German speaker male customers, for which we can diarize the opposite gender speakers by using inaSpeechSegmenter. Segmentation results can be seen in Figures \ref{fig:inaLabelsGerman1} and \ref{fig:inaLabelsGerman2}. We filter out non-speech segments using these labels and only consider speech segments shown as green in Figures \ref{fig:speechLabelsGerman1} and \ref{fig:speechLabelsGerman2}. We bundle each speaker's segments, and create two separate speaker streams from these bundles. We slide a 10 seconds wide window on this stream and by shifting it 1 second at a time, we created a data sample for each second of the speaker segments. 

In the second test, we preferred to use two sales call samples from a completely different source. We found two YouTube videos \cite{SCE:2012}, \cite{SCE:2014} of call center calls compatible with our problem definition. The calls are between English speaking female agents and customers.  After converting these videos to mp3 format, we processed them in the same way as the German test sample. 

Even the training is performed using Turkish audio samples; we see that the network can also generalize for German and English conversations. If we assume $y_i$ represents one of the classes, we want to find class probabilites $P(y_i|x_{j,S_l}^k)$ for each given utterance $x_{j,S_l}^k$. $i$ is either 0 or 1 for 2 Class Case and 0, 1, 2, or 3 for 4 Class Case with the labeling convention in Table \ref{tab:labelconv}. $k$ represents the actual label of the utterance, $j$ is the utterance number, and $S_l$ represents $l$th speaker, which is either 0 or 1 for the majority of call center calls. Since we want to predict the most probable class for a speaker ($S_l$), we first find the probability for a speaker to be in either class ($y_i$). We find the mean of the class probabilities for each utterance of the speaker as in Equation \ref{eq:classprob}. The ultimate label prediction can be predicted by using Equation \ref{eq:prediction}.  

\begin{align}
    P(y_i|S_l) &= \frac{1}{N_l}\sum\limits_{j=0}^{N_l} P(y_i|x_{j,S_l}^k) \label{eq:classprob}\\
    \hat{y}_{S_l} &= \argmax_{y_i} P(y_i|S_l)\label{eq:prediction}
\end{align}

We present the calculated class probabilities of the above tests in Table \ref{tab:gentest}. In the last column of this table, we show the calculated class probabilities. For example, for the German Sample 1, we see that female agent speaker audio achieves the maximum probability for class 2, considering  Table \ref{tab:labelconv}, representing ``female agent''. Hence, if we predict labels according to Equation \ref{eq:prediction}, all speakers are correctly classified with the correct labels.  We observed a 78\% match for the male customer in English Sample 1. When we closely analyze the segments detected by inaSpeechSegmenter, we observed that this slight inconsistency is due to faulty gender labels. These results support the idea that our network can generalize for the same problem in different languages. The network, with its optimized parameters, can work for other languages. It is also possible to fine-tune using transfer learning \cite{Yang:2020} with some training data from the desired language.

\begin{table*}[]
\caption{German - English Call Samples' Analysis Results. Maximum class probabilities are represented in bold.} 
\centering 
\begin{tabular}{l l l } 
\hline\hline 
 Example & Speaker & Class Probabilities ($P(y_i|S_l)$)\\ [0.5ex]
\hline 
 German Sample 1&female agent & [3.0491428e-08, 8.8888891e-02, \textbf{9.1111112e-01}, 4.6577166e-11] \\
 &male customer & [1.4224291e-13, \textbf{9.9999988e-01},  1.3527723e-07 1.6293862e-10]\\
\hline
 German Sample 2&female agent & [9.8589545e-09, 4.8513702e-08, \textbf{1.0000000e+00}, 1.4249832e-11]\\
 &male customer & [6.9357502e-12, \textbf{9.7998512e-01}, 2.0000000e-02, 1.4876848e-05]\\
\hline
 English Sample 1 \cite{SCE:2012} &female agent & [5.0117347e-14, 2.1594511e-11, \textbf{9.7777778e-01}, 2.2222223e-02]\\
 &male customer &[2.0569872e-29, \textbf{7.6325065e-01}, 2.5448738e-24 2.3674934e-01]\\
\hline
 English Sample 2 \cite{SCE:2014} &female agent & [3.3101374e-20, 4.8433914e-24, \textbf{1.0000000e+00}, 6.1925060e-35]\\
 &male customer &[6.5354442e-17, \textbf{9.9999976e-01}, 8.9157199e-17, 2.2629905e-07]\\
 \hline
\end{tabular}
\label{tab:gentest}
\end{table*}




\section{Concluding Remarks}

In this study, we presented a CRNN based solution to the speaker classification problem specific to call centers. A typical call center conversation takes place between a customer and an agent. Conventional call center software stores the calls with maximum possible compression rates as single-channel audio records. Hence, it is not possible to directly measure the individual speaker speech length for these records. If the speaker segments are a priori separated and fed to the proposed network, it can classify speakers into four classes, female customer, male customer, female agent, and male agent, with very high accuracy. 

We used the company's agents' gender information and the calls between two opposite genders to prepare the training data. Hence, we obtained a data set of utterances from 377 different speakers with known class labels. Thirty-eight speakers' utterances were never used in training and kept for validation. We also diversified the network structure by replacing GRUs with LSTM units in the recurrent layer. Training and validating with the data set yielded HLP comparable accuracies shown in Figures \ref{fig:gruresults} and \ref{fig:lstmresults} for both versions of the network. These results are summarized in Table \ref{tab:grulstmcomp}.

We also tested the network's generalization capability by using a couple of call samples from the German and English languages. Even without a thorough transfer learning, the trained network performed well on the selected samples. 
Since the study only focuses on the classification of speaker segments, we did not perform a thorough speaker separation or diarization process. With the current configuration, the proposed network can only analyze conversations between two opposite genders. Moreover, the system accuracy is inherently limited with the accuracy of inaSpeechSegmenter, which tends to mislabel speaker gender. Hence as a future study, we want to train another network for speaker separation for our single channel call center recordings, which can perform speaker separation regardless of gender and be used before the proposed network. Such a system can analyze the conversations between opposite genders and conversations between the same genders, i.e., female customer - female agent and male customer - male agent pairs. 
Calculating the individual speaker times gives a valuable metric to evaluate call center agents' performances. Also, unusual class-probability values may be interpreted to detect possible emotional abnormalities in the conversations between the customers and agents.

\bibliographystyle{spmpsci}      
\bibliography{references.bib}   

\end{document}